\documentclass[journal]{IEEEtran}
\usepackage{graphicx}
\usepackage{amsmath}
\usepackage{hyperref,graphicx}
\usepackage{color}
\usepackage{nopageno}
\usepackage{multicol}
\usepackage{nomencl}

\begin{document}
%\title{Investigation of Linear Properties of Electrostatic Wave in a Fermi Plasma}
\title{Linear Properties of Electrostatic Wave in Two-Component Fermi Plasma}

\author{Shraddha Sahana, Spandita Mitra, Swarniv Chandra and Suman Pramanick
\thanks{Shraddha Sahana is pursuing her B.Sc. degree from University of Calcutta, West Bengal 700073, India (E-mail: shraddhasahana10@gmail.com)}% <-this % stops a space
\thanks{Spandita Mitra is with Earth and Atmospheric Science, National Institute of Technology, Rourkela, Odisha 769001, India (E-mail: spanditamitra1998@gmail.com)}
\thanks{Swarniv Chandra is with the Physics Department, Government General Degree College at Kushmandi, Dakshin Dinajpur 733121, India (E-mail: swarniv147@gmail.com)}
\thanks{Suman Pramanick is with The Department of Physics, Indian Institute of Technology Kharagpur, Kharagpur, West Bengal 721302, India (E-mail: sumanhorse@iitkgp.ac.in)}

}
% <-this % stops a space
%\thanks{Manuscript received $<$26.8.2020$>$; revised $<$29.8.2020$>$.}
\pagenumbering{gobble}
\maketitle

\begin{abstract}

We study the two component Fermi plasma. Two components are electrons and ions. Using the Quantum-Hydrodynamic model (QHD), we study the linear properties of electrostatic wave. We derive the linear dispersion relation for the system from dynamical governing equations for the system. We study the dependence of linear-dispersion relation on various parameters of the system. \\

\end{abstract}

\section*{Nomenclature}
\hspace{-3.5mm}EW-Electrostatic Wave; QHD: Quantum Hydrodynamics;\\ DR- Dispersion Relation
\section{Introduction}
\IEEEPARstart{I}{}n recent times, scientists believe that the universe consists of 69 $\%$ of dark energy, 27$\%$ dark matter, and 1$\%$ normal matter, and all these celestial objects we see in the night sky are in the plasma state. Here, plasma (not the blood plasma) is an ionized gas that consists of electrons and ions.
Plasma is called ``The Fourth State of Matter" \cite{chen1984introduction}. Nowadays, Quantum plasma has become a major topic for research. Its application is spreading from neutron stars, giant planets, dwarfs to laboratory plasmas, also in electron-
hole plasma, electron gas etc. The basic condition for Quantum plasma is low temperature and sufficiently high density, unlike classical plasma \cite{pramanik2015international}. In extreme densities
when the distance between particles becomes as close as the quantum level, quantum tunneling plays a significant role. An electron feels a pressure due to its quantum effect. The examples of these pressures are Fermi pressure, relativistic pressure etc.\cite{goswami2020collision}. In Quantum plasma field, many researchers (Haas et. al, 2003 \cite{haas2003quantum}; Ali and Shukla, 2006 \cite{ali2006dust}; Manfredi, 2005 \cite{manfredi2005fields} ) have used Quantum Hydrodynamic Model (QHD). This QHD model treats plasma as a fluid and uses fluid equations for investigating this \cite{chandra2013electron}.

The electromagnetic fields in plasma have two parts- electrostatic and oscillatory. Waves are differentiated as electromagnetic or electrostatic according to there is any oscillatory magnetic field or not. And based on these, we can say, EWs must be longitudinal. So, If the magnetic field is zero then the wave is EW \cite{chen1984introduction}. Now we will talk about dispersion properties. We know that plasma is a dispersive medium. Dispersion relation (DR) describes the dispersive effects on the properties of waves in plasma. It relates the frequency of wave with its wave number. Whenever dispersion is present, wave velocity cannot be defined separately, we need to define group and phase velocity. The linear and non-linear properties of electron-acoustic solitary waves with three-component Fermi plasma has been studied in  \cite{pramanick2020electron}. For three-component Fermi plasma the Rouge-wave formation and dynamical properties of electron-acoustic solitary waves has been done in \cite{ghosh2020dynamical}.

The paper is organized in the following structure: In Sec-[I], we have discussed about plasma waves and DR. In Sec- [II],
we will set the governing equations using the QHD model and will normalize them with normalization schemes. In
Sec-[III], we will derive the dispersion relation using normalized equations and standard reductive perturbation method. In Sec- [IV], we will discuss about plots of DR. And then, we will conclude the effects of dispersion relation on 
electrostatic waves in Sec-[V].

%\subsubsection{Subsubsection Heading Here}
%Subsubsection text here.

\section{Basic Equations}\label{BE}
Let us consider the propagation of electrostatic waves in
Quantum plasma consisting of electrons and ions. Also,
assuming the plasma particles act like Fermi gas at zero temperature, the pressure term will be-
%\begin{equation}\label{eq1}
\begin{equation}
P_{j}=\frac{m_{j} V_{F j}^{2}}{3 n_{j 0}^{2}} n_{j}^{3}
\end{equation}
where, $j=e$ for electron \\ $j=i$ for ion \\ $m_{j}$= mass\\ $n_{j}$= Number density \\ $V_{F j}$=$\frac{\sqrt{2 k_{B} T_{F j}}}{m_j}$= Fermi speed \\  $T_{F j}$ = Fermi temperature\\$k_{B}$= Boltzmann's constant.\\
%\end{equation}
So, the set of QHD equations governing the dynamics of
Quantum plasma waves in a two-component plasma is listed
below-
\begin{equation}\label{eq2}
\frac{\partial n_{e}}{\partial t}+\frac{\partial\left(n_{e} u_{e}\right)}{\partial x}=0
\end{equation}
\begin{equation}\label{eq3}
\frac{\partial n_{i}}{\partial t}+\frac{\partial\left(n_{i} u_{i}\right)}{\partial x}=0
\end{equation}

\begin{equation}\label{eq4}
\left(\frac{\partial}{\partial t}+u_{i} \frac{\partial}{\partial x}\right) u_{i} =\frac{1}{m_{i}}\left[Q_i \frac{\partial \phi}{\partial x}+\eta_{i} \frac{\partial^{2} u_{i}}{\partial x^{2}}\right]
\end{equation}

\begin{equation}\label{eq5}
0=\frac{1}{m_{e}}\left[Q_e \frac{\partial \phi}{\partial x}-\frac{1}{n_{e}} \frac{\partial P_{e}}{\partial x}+\frac{\hbar^{2}}{2 m_{e}} \frac{\partial}{\partial x}\left[\frac{1}{\sqrt{n_{e}}} \frac{\partial^{2} \sqrt{n_{e}}}{\partial x^{2}}\right]\right]
\end{equation}
\begin{equation}
    \frac{\partial^{2} \phi}{\partial x^{2}}=4 \pi \left(Q_{e}n_{e}-Q_{i}n_{i}\right)
\end{equation}

where,\\
$u_j$= Fluid velocity\\
 $p_j$= Pressure\\
 $q_j$= Charge\\
 $j=e$ (For electron)\\
 $j=i$ (For ion)\\
Here, $q_e= -e$,  $q_i= +e$\\
$\hbar$=$(h / 2 \pi)$ where $h$ is the Planck's constant\\
\\
Now the normalization schemes for normalizing the
governing equations which we have used are-\\
$x\rightarrow x \omega_{pe} / V_{F e}, \quad  t \rightarrow t \omega_{pe},\quad \phi \rightarrow e \phi / 2 k_{B} T_{F e},\\ u_{j} \rightarrow u_{j} / V_{F e} , \quad n_{j} \rightarrow n_{j} / n_{j 0}$\\
where,
\begin{equation*}
\omega_{pe}=\sqrt{\frac{4 \pi n_{0} e^{2}}{m_{e}}}\text{is the electron plasma wave frequency}    
\end{equation*}
\begin{equation*}
 V_{Fe}=\sqrt{\frac{2k_{B}T_{Fe}}{m_e}}\text{ is Fermi speed}   
\end{equation*}
\begin{equation*}
H=\frac{\hbar \omega_{pe}}{2k_{B}T_{Fe}} \\    
\end{equation*}

Using the normalization consideration, equation (2) to
equation (6) can be rewritten as-

\begin{equation}\label{eq7}
\frac{\partial n_{e}}{\partial t}+\frac{\partial\left(n_{e} u_{e}\right)}{\partial x}=0
\end{equation}
\begin{equation}\label{eq8}
\frac{\partial n_{i}}{\partial t}+\frac{\partial\left(n_{i} u_{i}\right)}{\partial x}=0
\end{equation}

\begin{equation}\label{eq9}
\left(\frac{\partial}{\partial t}+u_{i} \frac{\partial}{\partial x}\right) u_{i}=-\mu\frac{\partial \phi}{\partial x}+\eta_{i} \frac{\partial^{2} u_{i}}{\partial x^{2}}
\end{equation}

\begin{equation}\label{eq10}
0=\frac{\partial \phi}{\partial x}-n_{e} \frac{\partial n_{e}}{\partial x}+\frac{H^{2}}{2} \frac{\partial}{\partial x}\left[\frac{1}{\sqrt{n_{e}}} \frac{\partial^{2} \sqrt{n_{e}}}{\partial x^{2}}\right]
\end{equation}

\begin{equation}\label{eq11}
\frac{\partial^{2}\phi}{\partial x^{2}}=\left(n_{c}-n_{i}\right)
\end{equation}

Here,
\begin{equation*}
    \mu=\frac{m_e}{m_i}
\end{equation*}
\begin{equation*}
    \eta=\frac{\eta_{i}V^{2}_{Fe}}{\omega_{pe}}
\end{equation*}
\begin{equation*}
    H=\frac{\hbar\omega_{pe}}{2k_{B}T_{Fe}}
\end{equation*}
H is a quantum parameter which is dimensionless and
proportional to quantum diffraction. $\hbar\omega_{pe}$is the energy of initial oscillation of plasma waves and $k_{B} T_{Fe}$ is the Fermi energy.

\section{Analytical Studies }\label{DR}
In order to study linear properties of plasma waves we use
the reductive perturbation expansion for the field quantities
$n_e$ , $n_i$ , $u_e$ , $u_i$ and $\phi$ about equilibrium values listed below-
\begin{equation}\label{eq12}
\left[\begin{array}{c}
n_{j} \\
u_{j} \\
\phi
\end{array}\right]=\left[\begin{array}{c}
1 \\
u_{0} \\
\phi_{0}
\end{array}\right]+\varepsilon\left[\begin{array}{c}
n_{j}^{(1)} \\
u_{j}^{(1)} \\
\phi^{(1)}
\end{array}\right]+\varepsilon^{2}\left[\begin{array}{c}
n_{j}^{(2)} \\
u_{j}^{(2)} \\
\phi^{(2)}
\end{array}\right]+\cdots
\end{equation}
Now substituting the expansion (12) in equations (7) to (11),and then linearizing and assuming all these field quantities changes like $e^{i(k x-\omega t)}$ , we got $k$ which is wave number and $\omega$ which is normalized wave frequency. So, the derived dispersion relation is-

\begin{equation}
    \omega=k u_{0}\pm k\sqrt{\frac{\mu\left(4+H^{2}k^2\right)}{k^2\left(4+H^2k^2\right)+4}}
\end{equation}
Where,$\quad \mu=\frac{m_e}{m_i},\quad H=\frac{\hbar\omega_{pe}}{2k_{B}T_{Fe}}$

 The equation (13) represents the dispersion relation of
electrostatic waves in Fermi plasma.
This quadratic equation has two solutions-
\begin{equation*}
    \omega=k u_{0}+ k\sqrt{\frac{\mu\left(4+H^{2}k^2\right)}{k^2\left(4+H^2k^2\right)+4}}
\end{equation*}
\begin{equation*}
    \omega=k u_{0}- k\sqrt{\frac{\mu\left(4+H^{2}k^2\right)}{k^2\left(4+H^2k^2\right)+4}}
\end{equation*}
During the calculation of dispersion relation, we have assumed the particles to be free from viscosity.

 \section{Results and Discussions}\label{res}
The linear dispersion relation of electrostatic wave in Fermi plasma with the
help of one dimensional QHD model and standard reductive
perturbation method has been investigated. 
Keeping all the
parameters within their range, we have plotted the dispersion
curve for different quantum diffraction parameters (H) and
different streaming velocities $u_0$. And we have got similar
type of two dispersion curves that have been plotted below.

\begin{figure}[!]
{	\centering
	\includegraphics[width=3.1in,angle=0]{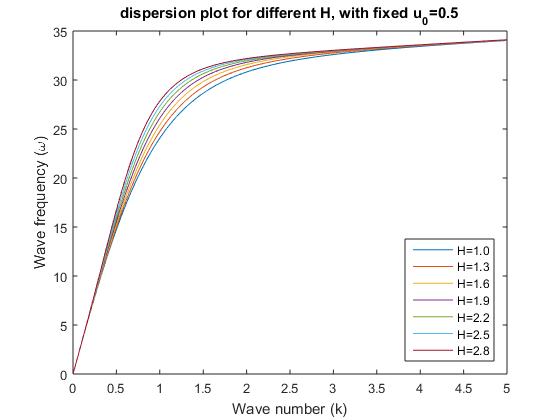}
	\caption{2D plot of dispersion relation for different Quantum diffraction
parameter (H) with $u_0$ = 0.5, $\mu$= 1000.}
	\label{fig1}
}
\end{figure}

\begin{figure}[!]
{	\centering
	\includegraphics[width=3in,angle=0]{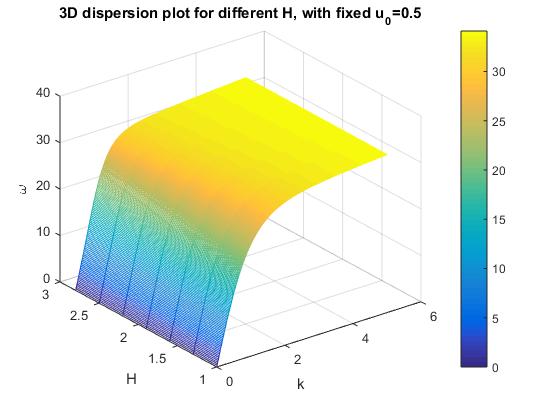}
	\caption{3D plot of dispersion relation for different Quantum diffraction parameter (H) with $u_0$ = 0.5, $\mu$= 1000.}
	\label{fig2}
}
\end{figure}

\begin{figure}[t]
{	\centering
	\includegraphics[width=3in,angle=0]{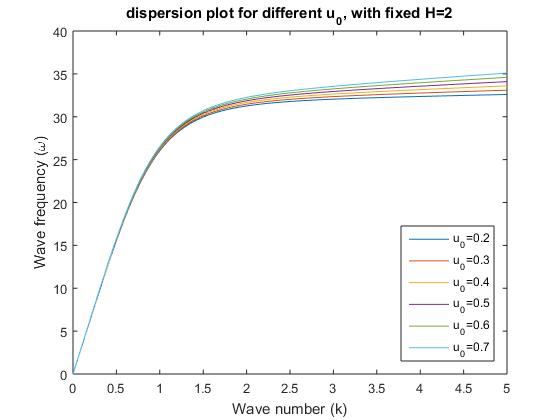}
	\caption{Dispersion relation 2D plot for different streaming velocities $u_0$ with H=2, $\mu$=1000.}
	\label{fig3}
}
\end{figure}

\begin{figure}[!t]
{	\centering
	\includegraphics[width=3in,angle=0]{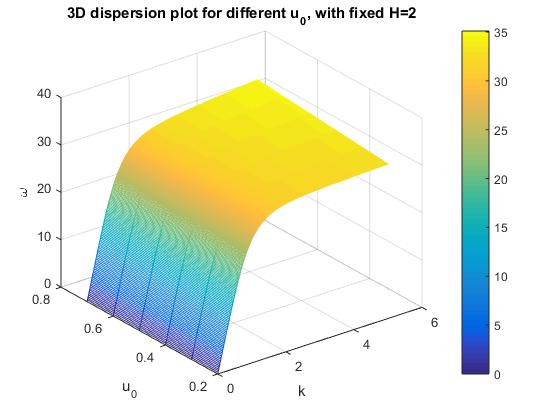}
	\caption{Dispersion relation 3D plot for different streaming
velocities $u_0$ with H=2, $\mu$=1000.}
	\label{fig4}
}
\end{figure}

The dispersion curve corresponds to electrostatic
waves, we have seen in both types of graphs keeping one
parameter constant and varying the other parameter, the
curves show the upward trend (We have plotted the
dispersion curve with positive $\omega$ vs $k$).

In fig (\ref{fig1}) and (\ref{fig2}), keeping the streaming velocity $u_0$
constant and increasing the values of H we have plotted the
graphs. An increase in H shows non-linearity in the graph.
And after some region, $\omega$ attains maximum value.
In fig (\ref{fig3}) and (\ref{fig4}), keeping H constant and increasing
the values of $u_0$ we have seen the shift towards the upper
direction in the wave frequency vs wave number graph.

\section{Conclusions}\label{Conc}

The Fermi-plasma consists of non-relativistic
electrons and ions. The dependency of wave frequency on
Quantum diffraction parameter and streaming velocity are
studied thoroughly. It is shown that these two components H
and $u_0$ have an important role in determining the linear
properties of electrostatic waves.
At high wave-number region, the dispersion curve with
constant H starts to split into different lines for different
streaming velocities $u_0$ because at the high wave-number
region the energy carrying capacity for different $u_0$ is
different.
In the graph of dispersion, with increasing H, the
quantum effect is increasing and so the non-linearity is
increasing as well. At high wave-number range, the
frequency becomes high, and for this reason, energy
becomes high. And because of the high energy, the curve
becomes observable in the classical range. That's why it is
classically linear everywhere. At a higher range of
wave-number, $\omega$ attains a maximum value.
The newly gotten results will be useful for
understanding the dispersion properties and obtaining the
group velocities and phase velocities of the waves. And it
will be also helpful for studying the instabilities of the waves
in Fermi plasma.

\bibliographystyle{IEEEtran}
\bibliography{main}

\end{document}